# Needed for completion of the human genome: hypothesis driven experiments and biologically realistic mathematical models

Roderic Guigó[1], Ewan Birney[2], Michael Brent[3], Emmanouil Dermitzakis[4,9], Lior Pachter[5], Hugues Roest Crollius[6], Victor Solovyev[7], Michael Q. Zhang[8]

1. Grup de Recerca en Informàtica Biomèdica,
   Institut Municipal d'Investigació Mèdica—Centre de Regulació Genòmica
   Universitat Pompeu Fabra, C/Dr. Aiguader 80, 08003 Barcelona,
   Catalonia, Spain
2. European Bioinformatics Insitute, Wellcome Trust Genome Campus
   Hinxton, Cambridge, CB10 1SD, United Kingdom
3. Laboratory for Computational Genomics, Campus Box 1045, Washington University
   One Brookings Drive, St. Louis, Missouri 63130
4. Department of Genetic Medicine and Development, University of Geneva Medical School,
   1211 Geneva, Switzerland
5. Department of Mathematics, University of California Berkeley
   970 Evans Hall, Berkeley, California 94720
6. Genoscope & CNRS UMR 8030
   2 rue Gaston Crémieux, 91057 Evry Cedex, France
7. Department of Computer Science, Royal Holloway,
    University of London
   Egham, TW20 0EX, United Kingdom

8. **Computational Biology and Bioinformatics, Cold Spring Harbour Laboratory, P.O. Box 100, 1 Bungtown Rd., Cold Spring Harbor, New York 11724**

9. **Present address: The Wellcome Trust Sanger Institute, Wellcome Trust Genome Campus, Hinxton, Cambridge, CB10 1SA, UK**

**Correspondence should be addressed to RG**

With the sponsorship of ``Fundació La Caixa'' we met in Barcelona, November 21st and 22nd, to analyze the reasons why, after the completion of the human genome sequence, the identification all protein coding genes and their variants remains a distant goal. This may came as a surprise to many, since we learn from the textbooks that the genetic code—the instructions by means of which the DNA sequence encodes the amino acid sequence of the proteins—was deciphered in the early 1960's. Nevertheless, the lack of an accurate and complete gene catalogue is still limiting the impact of the human genome sequence on biomedical research.

At the meeting there was consensus among us that since the publication in early 2001 of the draft sequence (Lander et al., 2001; Venter et al., 2001) a great deal of progress has been made towards the identification and characterization of the functional elements encoded in the human genome—protein coding genes in particular. The completion of the human genome sequence, the sequencing of cDNA libraries to higher accuracy (e.g. Mammalian Gene Collection), and the sequencing of additional vertebrate genomes such as that of the mouse (Waterston et al., 2002), has been complemented by significant improvements in automatic gene annotation programs and pipelines (as examples, see Guigó et al., 1992; Solovyev et al., 1995; Zhang, 1997; Birney and Durbin, 1997; Korf et al., 2001; Alexandersson et al., 2003; Parra et al., 2003; Birney et al., 2004 for programs and pipelines developed by us), and increased resources towards manual curation of finished human chromosomes (e.g the HAWK workshops, http://www.sanger.ac.uk/HGP/ havana/hawk.shtml).

And yet we also agreed that a number of important obstacles must still be overcome if we are to reach the goal of having a complete catalogue of the genes and their transcript variants in the human and other important model genomes:

1. increasingly aggressive sequencing of cDNA libraries appears to have reached a plateau and is yielding only a fraction (which could be small) of lowly or rarely expressed transcripts (Wang et al., 2000).

2. an important fraction of so-called full-length cDNA sequences may not in fact include the complete 5' end of the transcript, as recent 5' RACE experiments suggest (Suzuzi et al., 2002). The determination of the correct 5' end of a gene is essential to identify its promoter region—and thus to investigate the expression pattern of the gene.

3. it is still difficult to distinguish non-functional pseudo-genes from ``bona fide'' genes, in particular from short intronless genes. In fact, recent analysis suggests that we may have seriously underestimated the number of human pseudo-genes, which could be as high as the number of functional genes (Waterson et al., 2002).

4. fast evolving, human specific genes may be very difficult to detect by sequence similarity searches, because they lack obvious counterparts or homologues in other genomes.

5. there are no methods to predict the pattern of alternative transcript variation of human genes from primary sequence data. Even the current estimates of the incidence of the

phenomenon—based on partial EST sequences—are notably discrepant (Modrek and Lee, 2002).

6. statistical in nature, current computational methods are trained to identify genes with features—codon composition bias, splice site sequences, etc.—characteristic of the genes known so far, but they may not extrapolate well to the identification of genes where these features are weak, such as short intronless genes, or unusual, such as low expressed genes with anomalous codon bias or repetitive composition. In particular, these methods deal poorly with the exceptions—which in same cases, could be not as uncommon as we currently think—to the canonical rules defining eukaryotic genes. These include overlapping genes or genes within introns, non-canonical splice sites, and selenoproteins.

These issues limit our ability to identify protein coding genes in the human genome to such an extent that, during our meeting in Barcelona, we could only agree on their number within a margin of several thousands (see also Pennisi, 2003). At the root of these limitations lies our still incomplete knowledge of what defines a eukaryotic gene, and what are the mechanisms by which the sequence signals involved in gene specification are recognized and processed in the eukaryotic cell. The theoretical models on which gene finding methods are based reflect this partial knowledge, and are consequently over simplistic and inadequate, and may lead to a partial and biased view of the gene content in the human genome.

While the importance of theoretical models may have not been widely recognized in molecular biology, it is only through theory—through the formalizing of our understanding of the casual relationships between natural phenomena—that the transition from a descriptive to a predictive science is possible. Inaccurate theoretical models lead to inaccurate predictions. In this regard, we would like to stress that there is no such thing as purely experimental gene finding, and that underlying all gene annotation methods there is always a model of what a gene is. Even in the so-called ``experimental'' prediction methods, based on the sequencing of cDNA molecules, an implicit model of a gene is employed to infer, usually by computational means, the amino acid sequence of the protein product encoded by the gene and the associated exon boundaries in the genome sequence. This information is seldom determined by experimental means. According to most prevalent models, the amino acid sequence of the protein encoded by a gene is the translation (via the genetic code) of the longest open reading frame (ORF) in the cDNA sequence initiated by an ATG codon and terminated by one of the three stop codons: TAA, TAG, and TGA. Actually, none of these assumptions is always true. The first ATG within an ORF does not always initiate translation (Zhang, 1998), and in fact translation may not even begin with an ATG codon (Hann et al., 1988). Similarly, translation is not always terminated at a stop codon. In selenoproteins, for instance, at least one TGA codon is used instead to incorporate selenocysteine in the polypeptide chain. On the other hand, because of RNA editing, the synthesized amino acid sequence may not be identical to the one dictated by the genetic code on the primary cDNA sequence, and because of programmed translational frame-shifting, the synthesized protein product may not be encoded by a ``stricto sensu'' ORF in the cDNA sequence. We assume these phenomena to be rare, but because these assumptions are implicit in our gene models, we may have been seriously underestimating their incidence. Selenoproteins, for instance, are systematically mispredicted in the sequenced vertebrate genomes (Kryukov et al., 2003).

A cDNA sequence is not only the key to inferring the amino acid sequence of the encoded protein, but it also delineates the exonic structure and the boundaries of the gene via its mapping and alignment on the genomic DNA. Such information is a prerequisite to characterizing the pattern of exonic variation and identifying the promoters of a gene, which in turn help us to understand how the expression and the function of a gene are modulated in the cell. Current computational methods that map and align cDNAs on genomic DNA use gene models that do take into account the split nature of eukaryotic genes. These models are however still primitive and this leads – contrary to prevalent thinking – to much uncertainty in the identification of gene structures through cDNA mapping and alignments. First, the existence of recently duplicated paralogous genes and pseudo-genes means that a single cDNA sequence may map to multiple locations in the genome. Second, even if the location of the gene in the genome sequence has been unequivocally determined, uncertainty may remain in the delineation of the exon boundaries; indeed, often the nucleotides at the 5' end of one exon may also match the genome sequence at the 3' end of the downstream exon. Third, while the inclusion of constraints reflecting the occurrence of the canonical GT-AG dinucleotides at the intron boundaries  (and also of the AT-AC dinucleotides at the boundaries of U12 introns) in the gene model may help in resolving uncertainties, numerous exceptions have been reported to these canonical rules (Burset et al., 2001). Again, presumptions implicit in our working models of genes may lead to incorrect gene annotations, even in the presence of experimental data, and to the underestimation of the importance of certain biological phenomena. Examples abound in the history of molecular biology: alternative splicing of protein coding genes, for instance, once thought to be a rare exception, it is today considered a prevalent phenomenon in vertebrate genomes. We may be suffering from the well known fact in human psychology—whose impact in science is underscored by Frank Close in his book: "Too Hot to Handle: the Race for the Cold Fusion"—that people "become so committed to a preconceived belief in something that contrary information is ignored or reinterpreted to fit with the 'facts'" (Close, 1991)

It is, however, when the cDNA is lacking, and there is no "experimental" support for a gene, except perhaps for partial EST sequences, and close homologies to known amino acid sequences are also lacking, that the theoretical models of the gene are of paramount importance. Indeed, most our hopes for correctly annotating those genes not represented in sequenced cDNA libraries, possibly a substantial fraction of all human genes, rest on the faithfulness and comprehensive nature of these models.

Gene models underlying most modern computational gene prediction methods have a strong probabilistic component. Even though we are starting to recognize the intrinsic stochastic nature of the eukaryotic gene, it is not this stochastic nature that existing computational gene finders have attempted to model; in fact they almost systematically ignore it. Rather, the probabilistic models underlying computational gene prediction attempt to capture the characteristic statistical patterns in the genomic sequence induced by the presence of protein coding genes. These patterns are largely due to the uneven usage of amino acids in proteins, the uneven usage of synonymous codons for the same amino acid, and the local dependencies between amino acids in the protein sequence imposed by structural constrains. The programs predict a gene when the observed pattern in the genome sequence appears more likely than not to be caused by the existence of a protein coding gene. That is, computational programs detect genes mostly by the imprint they leave on the sequence—the consequence, but not the cause, of their existence. This is, of course, totally different from the mechanism by which the genome sequence is decoded in the eukaryotic cell to yield amino acid sequences. We don't generally believe that the cellular machinery is computing codon bias along the genome sequence to trigger and control the biochemical processes resulting in protein synthesis. We rather believe that the cellular machinery recognizes relatively few ``cis-signals'' in the primary DNA, or in the intermediate RNA sequences to trigger and control these processes: aside from related enhancers/silencers, these are promoter elements and termination signals during transcription, branch sites and splice sites during splicing, and initiation and termination codons during translation. Non-specialists may be surprised to learn that among the tens of thousands of parameters on which the programs depend—perhaps more the number of genes in the human genome!—very few attempt to model the biological mechanisms by means of which these sequence signals are recognized during these processes.

Given the difficulties of the task (the protein coding fraction may be lower than 1.5% of the human genome), current computational gene predictions methods perform remarkably well. However, they are not accurate enough to produce reliable automatic annotations of the eukaryotic genomes. Thus, in the absence of a full length cDNA, our current gene computational predictions are hypotheses, that require experimental verification. Indeed, we believe it is important to stress that the current gene annotation of the human genome has a strong hypothetical component. Failing to recognize this may lead to substantial resources being wasted—in particular in small experimental

laboratories—for instance, by unsuccessfully attempting to amplify by RT-PCR a computational gene prediction that, even though corresponding to a real gene, had the intron boundaries mispredicted.

At the meeting in Barcelona we agreed that, to address the limitations of current computational gene finders, a strong shift is required in the nature of the models that we employ in them. Indeed, we believe that the mathematical models of the eukaryotic gene should incorporate a faithful formalization of the biological processes involved in gene specification. Modelling of splicing is a case in point. Even the most sophisticated computational models of splicing currently available are limited to model dependencies between positions within the canonical signals defining the intron boundaries. The models implicitly assume, in consequence, the splice signals to be recognized independently and atemporally in a nucleic acid sequence void of further information. There is, however, increasing experimental evidence suggesting that additional intronic and exonic sequences play a role in the definition of the intron boundaries, and in the regulation of the production of alternative splice forms. Moreover, there is a dynamic interplay—not yet completely understood—between transcription and splicing, and dependencies between distant splice signals can not be ruled out. RNA structure, as well, may influence splice site selection All these phenomena should be taken into account in a biologically realistic model of the splicing process. And, while experimental data is crucial to understanding the mechanistic details of these phenomena, the fact that we have accumulated in sequence databases a large collection of known instances of splicing makes the contribution of computational analysis very important.

The concept of gene is central to Biology—as central as, for instance, the concept of atom is in Chemistry. Its mathematical formulation will thus play a fundamental, almost founding role, in the edification of the theoretical framework of the molecular biology of the XXIst century. A theory of living systems—often referred to as Systems Biology—that will make possible the transition from a mostly descriptive to a highly predictive Biology.

"Without a theory, data would fly by unnoticed," writes William Gough in his review of Close's book (Gough, 1992). Indeed, the lack of an appropriate theoretical framework difficults the translating of the current flood of genomic data into relevant biological knowledge. The emergence of high-throughput techniques, characteristic of genomics research, has lead to the so-called data- or discovery-driven biology, in which data is obtained without the need for a hypothesis about the nature of biological phenomena, in contra-position to the classical hypothesis-driven approach in which experiments are performed (and data obtained) to test previously formulated hypothesis within the framework of a pre-existing theory. Genome Projects are mostly high throughput biology, and they certainly produce a lot of valuable data. High throughput biology alone, however—either through indiscriminate sequencing of cDNA libraries, or through genome wide expression tiling microarrays—appears to have reached a limit in its ability to annotate the genes in the human genome. For instance, we now start to see regions of the genome that are transcribed but do not appear to be coding for proteins. It is therefore time for the computational biologists to take up to the task of developing a powerful theory of the eukaryotic gene. A theory that would lead to improved computational gene predictions, and that would in particular resolve the apparent discrepancy between transcriptional surveys and the estimated protein coding density of the human genome.

In this regard, recent reports underscore the importance of a hybrid approach, in which high throughput biology is driven by computational predictions of hypothetical genes. In one such report, Guigó et al. (2003) computational gene predictions corresponding to real genes have been efficiently discriminated from those likely to be false positives by using a more complex model of the human genes that captures the extraordinary conservation of the exonic structure between human and mouse orthologous genes. Application of this method followed by experimental verification by RT-PCR has led to the identification of hundreds of novel human genes. Similarly, coupling of computational gene predictions with microarray profiling has recently suggested the existence of at least 2000 novel genes in the *Drosophila melanogaster* genome (Hild et al., 2003)—one of the best annotated of all higher eukaryotic genomes. In yet another example, experimental cloning of predicted genes in *Caenorhabditis elegans* lead to the corrections of the predicted exonic structure in 50% of the cases (Reboul et al., 2003).

In any case, while we disagreed on the relative value of high throughput experimental verification and manual curation, versus computational predictions in the annotation of the human genome, we concurred that only through a biologically realistic mathematical model of the eukaryotic gene—which would render both experimental verification and manual curation less necessary—we can hope to characterize, even if only approximately, the gene complement of the hundreds of genomes to be sequenced in the coming years, and for which the amount of resources we have devoted to the human genome will certainly not be available.